\documentstyle[12pt,epsf]{article}
\renewcommand{\baselinestretch}{1.45}

\date{\today}

\def\be{\begin{equation}}
\def\ee{\end{equation}}
\def\bear{\begin{eqnarray}}
\def\eear{\end{eqnarray}}
\def\nn{\nonumber}

\newcommand\px[1]{{\partial_{#1}}}
\newcommand\qx[1]{{\partial^{#1}}}

\newcommand\rep[1]{{\bf {#1}}}      % representation
\newcommand\tr[1]{{\mbox{tr}\{{#1}\}}}          % trace
\newcommand\com[2]{{\lbrack {#1},{#2}\rbrack}}  % commutator
\newcommand\MR[1]{{{\bf R}^{#1}}}               % Real numbers
\newcommand\MS[1]{{{\bf S}^{#1}}}               % Circle, sphere,...
\newcommand\AdS[1]{{{\bf AdS}_{#1}}}            % Anti-DeSitter
\newcommand\MT[1]{{{\bf T}^{#1}}}               % Torus
\newcommand\SUSY[1]{{{\cal N}= {#1}}}           % N=? SUSY
\def\a{{\alpha}}

\def\wR{{\widetilde{R}}}

\def\wstar{{\tilde{\star}}}       % dipole-product
\def\a{{\alpha}}                 % spinor index
\def\da{{\dot{\alpha}}}          % dotted spinor index

\newcommand\putfig[3]{
   \framebox[\textwidth]{
   \vbox{
   \vspace{0.5cm}
   \let\picnaturalsize=N
   \def\picsize{#3}
   \def\picfilename{#1}
   \ifx\nopictures Y\else{\ifx\epsfloaded Y\else\input epsf \fi
   \let\epsfloaded=Y
   \centerline{\ifx\picnaturalsize N\epsfxsize \picsize\fi
   \epsfbox{\picfilename}}}\fi
   \vspace{0.75cm}
   {\hspace{0.25truein}\parbox{5.5truein}%
{\renewcommand{\baselinestretch}{1} \small #2}}
   \vspace{0.5cm}
   }}
}  

\begin{document}

%%%%%%%%%%%%%%%%%%%%%%%%%%%%%%%%%%%%%%%%%%%%%%%%%%%%%%%%%%%%%%%%%%%%%%%%%%%%
%                    TITLE PAGE                                            %
%%%%%%%%%%%%%%%%%%%%%%%%%%%%%%%%%%%%%%%%%%%%%%%%%%%%%%%%%%%%%%%%%%%%%%%%%%%%

\begin{titlepage}
\titlepage
\rightline{hep-th/0010072, IAS-HEP-00/78, PUPT-1959}
\rightline{\today}
\vskip 1cm
\centerline{{\Huge Vector Deformations}}
\centerline{{\Huge of $\SUSY{4}$ Super-Yang-Mills Theory,}}
\centerline{{\Huge Pinned Branes, and Arched Strings}}
\vskip 1cm
\centerline{
Keshav Dasgupta$^\clubsuit$,
Ori J. Ganor$^\diamondsuit$ and
Govindan Rajesh$^\clubsuit$
}
\vskip 0.5cm

\begin{center}
\em  ${}^\clubsuit$School of Natural Sciences, \\
Institute of Advanced Study, \\
Einstein Drive, \\
Princeton, NJ 08540, USA \\
{\tt keshav, rajesh@ias.edu} \\
\end{center}

\vskip 0.5cm

\begin{center}
\em  ${}^\diamondsuit$Department of Physics, Jadwin Hall \\
Princeton University \\
Princeton, NJ 08544, USA \\
{\tt origa@viper.princeton.edu}\\
\end{center}

\vskip 0.5cm

\abstract{
We study a deformation of $\SUSY{4}$ Super-Yang-Mills theory
by a dimension-5 vector operator. There is a simple nonlocal
``dipole'' field-theory that realizes this deformation.
We present evidence that this theory is realized in 
the setting of ``pinned-branes.'' 
The dipoles correspond to open strings that arch out of the brane.
We find the gravitational dual of the theory at large $N$.
We also discuss the generalization to the $(2,0)$ theory.}
\end{titlepage}
             
%%% ------------------------- CUT HERE ---------------------------------%
%\vskip 0.5cm
%{\bf this is file ``intro.tex''}
%\vskip 0.5cm

\def\th{{\theta}} % noncommutativity parameter
\def\cO{{\cal O}} % operator
\def\cwO{{\widetilde{\cal O}}} % operator
\def\Lag{{\cal L}} % Lagrangian

\section{Introduction}\label{intro}
% =======================================================================
Consider the superconformal
$\SUSY{4}$ Super-Yang-Mills on a noncommutative $\MR{4}$
(For recent developments in field-theory on noncommutative spaces
see \cite{CDS,DH,SWNCG} and references therein).
We will refer to it as $\SUSY{4}$ NCSYM.
Let the noncommutativity be specified by a 2-form $\th^{ij}$
such that the commutator of the coordinates
on $\MR{4}$ is $\com{x^i}{x^j} = i\th^{ij}$.
The parameter $\th^{ij}$ has dimensions of ${\mbox{Mass}}^{-2}$.
At low-energies, NCSYM can be described by augmenting the action with:
$$
\int\th^{ij}\cO_{ij}(x) d^4 x,
$$
where $\cO_{ij}$ is an operator of dimension $6$ in the superconformal
SYM on a commutative space. 
In the conventions such that the SYM Lagrangian is:
\bear
\Lag_{SYM} &=& \tr{
\frac{1}{2g^2}\sum_{I=1}^6\px{i}\phi^I\qx{i}\phi^I
+\frac{1}{4g^2}F_{ij}F^{ij}
  +\frac{1}{2g^2}\sum_{I<J}\com{\phi^I}{\phi^J}^2}
  + {\mbox{fermions}},
\nn
\eear
the bosonic part of the operator $\cO_{ij}$ can be written as:
\bear
\tr{
\frac{1}{2g^2}F_{jk}F^{kl}F_{li}
-\frac{1}{2g^2}F_{ij}F^{kl}F_{kl}
+\frac{1}{g^2}F_{ik}\sum_{I=1}^6\px{j}\phi^I\qx{k}\phi^I
-\frac{1}{4g^2}F_{ij}\sum_{I=1}^6\px{k}\phi^I\qx{k}\phi^I},
\nn
\eear
Here, $g$ is the SYM coupling constant, $F_{ij}$ is the $U(N)$
field-strength, and $\phi^I$ ($I=1\dots 6$) are the scalars.

The operator $\cO_{ij}$ is neutral
under the $SU(4)$ R-symmetry and is part of a short representation
of the supersymmetry algebra. It is the same operator that
also describes the deformation of $\SUSY{4}$ SYM theory into
the Born-Infeld theory, to first order in the NSNS 2-form $B$-field
\cite{FLZ}.
The multiplet of operators to which $\cO_{ij}$ belongs also
contains two vector operators in the representation $\rep{15}$
of $SU(4)$, a self-dual tensor in the representation $\rep{10}$,
an anti-self-dual tensor in the representation
$\rep{\overline{10}}$ as well as several fermionic operators
(see for instance \cite{FFZ,AndFer,IntBON}).
In this paper we will concentrate on one of the 
vector operators, $\cO_i$ and $\cwO_i$.

Since the deformation by $\cO_{ij}$ can be extended to a complete
theory with a simple description, i.e. NCSYM, one might expect
that the deformation by $\cO_i$ can be interpreted as the 
first term in a low-energy expansion of a simple theory too.
This is indeed the case. The deformation by $L^i\cO_i$ (where
$L^i$ is a constant vector) is the low-energy expansion of
a nonlocal field-theory, the ``dipole-theory,'' described in \cite{BG}.
On $\MR{3,1}$, the non-perturbative description of NCSYM is more
subtle---in particular, it is known that S-duality of non-commutative
theories is problematic \cite{SSTii}-\cite{GMMS}.
This has to do with the appearance of time-like noncommutativity
\cite{SSTi}. The complete theory, NCOS, was described in \cite{SSTii, GMMS}.
These issues are further discussed in \cite{BarRab}-\cite{AhGoMe}.
It is likely that the dipole-theories need an extension as well,
especially if $L_i$ is time-like.

For simplicity, we can assume a Euclidean space.
The bosonic part of the $\SUSY{4}$ SYM operator $\cO_i$
can be calculated by changing to local variables (see \cite{BG}
for more details).
We can write it in $\SUSY{1}$ superfield notation as:
\bear
\cO_i &=& 
  {i\over {g_{YM}^2}}\int d^2\theta \epsilon^{ab}
   \tr{\sigma_i^{\a\da} W_\a \Phi_a D_\da\Phi_b +\Phi \Phi_a D_i\Phi_b}
   + {\mbox{c.c.}}
\label{cOvec}
\eear
Here, the $\SUSY{2}$ vector-multiplet was decomposed into an
$\SUSY{1}$ chiral field, $\Phi$, and an $\SUSY{1}$ 
vector-multiplet whose field-strength is $W_\a$.
The hyper-multiplet was decomposed into two chiral multiplets
$\Phi_a$ ($a=1,2$) and $\sigma_i^{\a\da}$ are Pauli matrices).
Note that $\cO_i$ has conformal dimension 5.

On the other hand, nonlocal theories that are parameterized by
a vector have been argued to appear on ``pinned-branes'' \cite{CDGG}.
In this construction, the low-energy SYM that appears 
in the low-energy limit
on branes  is modified when they are put in an NSNS 3-form
field-strength (see also \cite{PolStr} for related constructions).
This effect can be realized by placing the branes
near the center of a Taub-NUT
space and turning on an NSNS 2-form flux at infinity.
The NSNS flux has one
direction along the Taub-NUT circle and another direction along the brane.

The purpose of the present work is to connect the pinned-brane theories
to the dipole-theories. We will bring several pieces of evidence to 
indicate that they are one and the same theory:

\begin{itemize}
\item
Both theories preserve the same amount of supersymmetry and 
they break the same part of the Lorentz group.

\item
We will expand the IR part of the pinned-brane gravitational solution
and show that it is a perturbation of $\AdS{5}\times \MS{5}$ by $\cO_i$,
in the context of the AdS/CFT correspondence \cite{M,GKP,WAdS}.
This will also allow us to find the gravitational dual of
the dipole-theories at large $N$ analogous to the gravitational dual
of the noncommutative gauge theories \cite{HasItz,MalRus}.

\item
We will construct the dipoles as curved open strings that
extend out of the brane and rotate in the directions transverse
to the brane. They are stabilized by magnetic forces similar to
those discussed in \cite{Myers,MST}.

\item
We will compactify on $\MT{2}$,
turn on a magnetic field on the brane and show, from a BPS analysis,
that the transverse fluctuations become massive.
This agrees with the behavior of dipoles in a magnetic field
and the conjecture that the fields
that describe transverse fluctuations are dipoles.
\end{itemize}

The paper is organized as follows.
Section (\ref{dipole}) is a review of the dipole-theories.
In section (\ref{pinned}) we review the pinned-brane constructions
and expand the IR region as a perturbation of $\AdS{5}\times \MS{5}$.
In this section we also present our conjecture for the gravitational
dual of the dipole-theories.
In section (\ref{arch}) we construct the dipoles as open strings
that arch out of the brane and are held by (generalized) magnetic forces.
In section (\ref{bps}) we compactify on $\MT{2}$ and
study the behavior of the dipoles in a magnetic flux.
In section (\ref{twoz}) we present the generalization to the $(2,0)$ 
theory. We propose that there is a deformation of the $(2,0)$-theory
to a theory that contains {\bf discpoles} --
membrane-like objects that generalize the dipoles.

%%% ------------------------- CUT HERE ----------------------------------

\section{Dipole Theories}\label{dipole}
% =======================================================================
Dipole-theories can be thought of as a generalization of field
theories on commutative or noncommutative spaces.
They are constructed by modifying the ordinary (or noncommutative)
product of functions to the  $\wstar$-product defined as follows.
To each field, $\Phi(x)$, we assign a dipole-vector, $L^i$.
The complex conjugate field, $\Phi^\dagger(x)$ is assigned the 
dipole-vector $-L^i$. 
If $\Phi_1(x)$ and $\Phi_2(x)$ have dipole-vectors $L_1$ and $L_2$
respectively, we define their dipole-product to be \cite{BG}:
\be\label{defwst}
(\Phi_1\wstar\Phi_2)_{(x)} \equiv
   \Phi_1(x-\frac{L_2}{2})\Phi_2(x+\frac{L_1}{2}).
\ee
For associativity,
we have to make sure that the dipole-vector
is additive, i.e. that $\Phi_1\wstar\Phi_2$ is defined to have
dipole-vector $L_1+L_2$.
Intuitively, the dipole field $\Phi(x)$ represents a dipole
of length $L$ starting at the point $x-\frac{L}{2}$ and ending at
$x+\frac{L}{2}$.
To see this, let us add a $U(1)$ gauge field, $A(x)$, and define
it to have dipole-vector zero.
The covariant derivative is then:
$$
D_i\Phi(x) = 
\px{i}\Phi(x) -i A_i(x)\wstar\Phi(x) 
  + i\Phi(x)\wstar A_i(x)
=\px{i}\Phi(x) -i A_i(x-\frac{L}{2})\Phi(x) + i\Phi(x)A_i(x+\frac{L}{2}).
$$
So $\Phi(x)$ transforms nontrivially under the
subgroup $U(1)_{(x-\frac{L}{2})}\times U(1)_{(x+\frac{L}{2})}$
of the gauge group,
where $U(1)_{(x)}$ is the local transformation group at $x$.

To ensure associativity we can start with an ordinary theory
that has a global $U(1)$ symmetry and assign to the field $\Phi_I$
($I$ is an arbitrary index that labels the field)
a dipole-vector
in the form $L_I = Q_I L$ where $L$ is a fixed vector, common to
all the fields, and $Q_I$ is the $U(1)$ charge of $\Phi_I$.
More generally, working on $\MR{d}$,
we can have global charges, $Q_{Ia}$, $a=1\dots l$
and a fixed $d\times l$ matrix, $\Theta^{ia}$ ($i=1\dots d$), such that 
the field $\Phi_I$ is assigned a dipole-vector $\sum_a\Theta^{ia}Q_{Ia}$.

The claim that field-theories on a noncommutative space are a special
case of the dipole-theories can be interpreted in two ways.
First, the $\wstar$-product can be defined to modify
a noncommutative $\star$-product. We just have to interpret
the products on the RHS of  (\ref{defwst}) as $\star$-products.
Moreover, starting with a commutative space, we can take the
charges $Q_{Ia}$ above to be the components of the
momentum of the field $\Phi_I$
and then $a=1\dots d$. If $\Theta^{ia}$ is chosen to be anti-symmetric
we recover the familiar field-theory on a noncommutative $\MR{d}$.
Let us also note that gauge theories on a noncommutative space
can be recast in terms of bi-local fields \cite{IKK} which
might be reminiscent of the dipole-fields.\footnote{We are grateful
to N. Seiberg for pointing this reference out.}

In this paper we will concentrate on a dipole deformation of
$\SUSY{4}$ Super-Yang-Mills (SYM) in 4D.
The dipole-vectors will be correlated with a single $U(1)$ charge,
i.e. they are of the form $L_I = Q_I L$. The $U(1)$ global 
symmetry is chosen as a subgroup of the R-symmetry, $SU(4)$,
that breaks it to $SU(2)\times U(1)$ and preserves $\SUSY{2}$ 
supersymmetry. If we decompose the $\SUSY{4}$ gauge field multiplet
under $\SUSY{2}$ supersymmetry, we get a vector-multiplet, $V$, and
a hyper-multiplet, $H$. We take all the fields in the vector-multiplet
to have dipole-vector zero and all the fields in the hypermultiplet
to have dipole-vector $L$ (their complex conjugate fields will have
dipole vector $-L$).
The Lagrangian is obtained from the $\SUSY{4}$ Lagrangian
by modifying the product to the $\wstar$-product, in a way
that is similar to the construction of $\SUSY{4}$ SYM on a noncommutative
$\MR{4}$. We can take the gauge group to be either $U(n)$ or 
$SU(n)$. This is unlike SYM on a noncommutative $\MR{4}$, where the $SU(n)$
theory is not well defined because two $SU(n)$ gauge transformations
can close to a $U(n)$ gauge transformation when the
product is changed to the $\star$-product \cite{SWNCG}.
In our case, the gauge-fields have dipole-vector zero and the gauge group
is unmodified.

In the limit $L\rightarrow 0$, this particular dipole theory
can be recast as a small deformation of ordinary $\SUSY{4}$ SYM.
The deformation operator is of the form $\int L^i \cO_i(x) d^4 x$
where $\cO_i$ is the dimension five operator described in the introduction.

The dipole-theories can be realized as brane configurations
using a construction similar to that of \cite{WitBR}.
Take an NS5-brane in directions $0\dots 5$ and compactify the
$6^{th}$ direction on a circle of radius $R$ but
such that the identification is: $(x_1,x_6)\sim (x_1 + L,x_6+2\pi R)$.
Now take a D4-brane in directions $0\dots 3,6$.
We are considering the limit $M_s^{-1}\ll R\ll L$ (where $M_s$
is the string-scale).
In this construction it is reasonable to expect that the 
open strings that connect the D4-branes on two sides of the NS5-brane
will become dipoles.

In order to find the gravitational dual of the theory and
especially in order to study the more interesting extension
to the $(2,0)$ theory it will be useful to study another 
realization of the dipole theories where the branes on which
the dipoles ``live'' are D3-branes. For that purpose we
T-dualize along the $6^{th}$ direction to turn the D4-brane 
into a D3-brane and the NS5-brane into a Taub-NUT space.

\section{Pinned branes}\label{pinned}
% =======================================================================
In \cite{CDGG}, a class of non-Lorentz invariant field-theories
was constructed by placing D$p$-branes, M2-branes or M5-branes
in a Taub-NUT space with NSNS or 3-form flux turned on at infinity.
Specifically, consider type-II string-theory and take the four
dimensional Taub-NUT solution:
\be\label{tnmet}
ds^2 = R^2 U(dy - A_i dx^i)^2 + U^{-1} (d\vec{x})^2,\qquad
i=1\dots 3,\qquad 0\le y\le 2\pi.
\ee
where,
$$
U = \left(1 + {{R}\over {|\vec{x}|}}\right)^{-1},
$$
and $A_i$ is the gauge field of a monopole centered at the origin.
The Taub-NUT solution is a fibration of a circle
over $\MR{3}$ such that at $|x|\rightarrow\infty$ the circle 
has a constant radius, $R$.
We now let $n$ D$p$-branes probe this geometry.
The D$p$-branes are points in the Taub-NUT space and
extend along $(p+1)$ of the other 6 directions. We assume $p\le 4$.
At infinity, we set the boundary condition on the NSNS 2-form
B-field to approach a nonzero 2-form constant, $b_{ij}$.
The 2-form is taken
to have one direction along the Taub-NUT circle, and the other direction
can be taken either parallel or transverse to the D$p$-brane.

This configuration preserves 8 supersymmetries (i.e. $\SUSY{1}$
for D4-branes, $\SUSY{2}$ for D3-branes, $\SUSY{4}$ for D2-branes
and so on).

Let us first consider the case in which the $B$-field is set transverse 
to the D$p$-brane. To be specific, let the D3 brane be oriented along
$x^0, x^1,.., x^3$.
We keep it fixed at the origin of a Taub-NUT space whose
nontrivial metric is along $x^6, x^7, ..., x^9$. The Taub-NUT circle
is $x^6$. Using the results of \cite{CDGG} we can show that the $B$ field
as seen by the D3 brane is 
$$
B=h~tan~\theta~dx^5\wedge(dx^6+A_idx^i)
$$
where $i=7,8,9$ and $B(r\to \infty)=tan~\theta \equiv b$ is the value of $B$
field at $r=\sqrt{x^ix^i}=\infty$. We also define
$$
h^{-1}= sin^2\theta +\left(1+{R\over r}\right)cos^2\theta
$$
The string coupling is
$g=e^{\phi}=\sqrt{h
\left(1+{R\over r}\right)
}$. Defining $G_{ij}$ as
the metric for the system, we can use a similar idea as in \cite{CDGG} to show
that the D3 brane is ``pinned''. The pinning potential for this case will be
$$
{\sqrt{detG}\over g}= cos~\theta = {1\over \sqrt{1+b^2}}
$$
The low-energy description of the D3-branes is $U(n)$ SYM
with a massive adjoint hyper-multiplet. The mass is given by:
$$
m^2 = {b^2\over 1+b^2}
$$
On the other hand, if the $B$-field with one leg
is set parallel to the D3-branes,
we get a nonlocal $(p+1)$-dimensional theory that is a deformation
of SYM.
To study the IR limit of the theory for a D3-brane we have to determine the 
low energy supergravity solution of the background. 
Let us take the direction of the $B$-field along the D3-brane
to be the $1^{st}$. Solving the equations of background supergravity
 we can show that
the component of $B$ field parallel to the D3 brane is given by
$$B_{16}= h~tan~\theta$$
where $h$ is the same function as before.
The behavior of the dilaton or the
string coupling is again identical to the previous case.
However the pinning potential is now
$$
{\sqrt{detG}\over g}= 1
$$
therefore there is no pinning! This is a generic phenomenon
 for branes in the 
background of Taub-NUT and $B$-fields with one leg parallel to the branes.
The other leg of the $B$-field should be along the compact $x^6$ circle of
Taub-NUT. The existence of Taub-NUT therefore reduces the worldvolume
supersymmetry to ${\cal N}=2$ and also ensures that we cannot gauge away the
$B$-field.

In the large $n$ limit we have a large number of D3 branes near a Taub-NUT
singularity. However the point $r\to 0$ is a coordinate singularity
and in the right choice of coordinate system $r=u^2$ Taub-NUT is actually
a smooth manifold. Therefore the world volume theory of the D3 brane is a
deformation of the ${\cal N} =4$ supersymmetry which preserves ${\cal N} =2$
supersymmetry. This deformation is due to the vector $B_{1i}$ and it creates
a scale in the theory. This scale is the dipole length and therefore from the
supergravity point of view we have a deformed $\AdS{5}$. In the rest of this
section we will elaborate on this issue. 

For our configuration, in the absence of $B$ field,
 the metric has two components $ds^2_{\parallel}=ds^2_{0123}$ and
$ds^2_{\perp}=ds^2_{45}+ds^2_{Taub-NUT}$. The AdS background is then given by 
$$ds^2= H^{-1/2}ds^2_{\parallel}+H^{1/2}ds^2_{\perp}$$ where $H$ is the 
harmonic function of the D3 branes. When we switch on a $B$ field such that
its asymptotic value is small the background metric gets deformed,
for small $r$, to:
$$ds^2_{\parallel}\to ds^2_{\parallel} -rb^2(dx^1)^2,~~~
ds^2_{\perp}\to ds^2_{\perp} + O(r^2)(dx^6+A_idx^i)^2$$

Therefore near $r=0$, the metric 
is $\AdS{5}\times\MS{5}$ and the NSNS B-field is $B_{16}= b~r$. 

Thus, this space is a deformation of the $\AdS{5}\times\MS{5}$ solution
and the deformation approaches zero as $r\rightarrow 0$.
In the AdS/CFT correspondence \cite{M,GKP,WAdS}, such
a deformation corresponds to a deformation of $\SUSY{4}$ SYM
by an irrelevant operator.
Recall that a field that behaves as $r^\delta$ corresponds to
a deformation $\int \cO(x) d^4 x$, where the operator $\cO$
has dimension $4+\delta$. In our case 
$\delta=1$ and this operator is the same as the
one in equation (\ref{cOvec}).
In fact, in \cite{FFZ,AndFer,IntBON} 
the list of operators in $\SUSY{4}$
SYM that belong to short SUSY representations was calculated from
the AdS/CFT correspondence. It was found that there are two
operators that are  vectors and are descendants of the chiral
primary $\tr{\Phi^{(I_1}\Phi^{I_2}\Phi^{I_3)}}$ where
$\Phi^I$ are the scalars of $\SUSY{4}$ SYM ($I=1\dots 6$) and 
$(I_1 I_2 I_3)$ denotes complete symmetrization.
These vector operators are in the representation $\rep{15}$ of the R-symmetry
group $SU(4)$. They correspond to (\ref{cOvec}) and its magnetic dual.
Thus, we have supporting evidence to the claim that the 
(un)pinned-branes in this case realize the dipole-theories.

\section{The dipoles as arched strings}\label{arch}
% =======================================================================
Consider the background of a Taub-NUT space before
we have placed the D3-branes.
In section (\ref{pinned}) we have found the geometry of
the Taub-NUT space with the $B$-field turned on at infinity.
After the appropriate change of variables, $r = u^2$, it is easy to
see that the origin, $r=0$, is nonsingular and the 3-form NSNS
flux $H = dB$ has a finite magnitude. 

To complete the identification of the theory on the D3-brane in the above
background with the dipole theory, we need to interpret the
quanta of the dipole fields that arise on the brane.
Clearly, the only objects that are charged under the $U(1)$ of the
D3-brane are open strings. However, open strings would seem to shrink to zero
size by their own tension.

We propose the following resolution to this puzzle.
Instead of considering a quantum of the dipole-field
that has one unit of R-symmetry charge, let us
consider a classical object with a large amount of R-symmetry
charge.
It is sufficient to restrict to the vicinity of the origin of
the Taub-NUT space, so
let us take a D3-brane in directions $0,1,2,3$ and an NSNS
3-form $H$-flux in directions $1,6,7$.
We will now construct an object with a large amount of angular momentum
in the $6,7$ plane that behaves like a dipole.
In principle, R-symmetry corresponds to a simultaneous rotation
in the $6-7$ and in the $8-9$ plane by the same angle.
Thus, the object is formed by an open string that is parameterized as:
$$
x_0 = \tau,\,\,
x_1 = \frac{L}{\pi}\sigma,\,\,
x_2 = x_3 = x_4 = x_5 = 0,\,\,
x_6+ i x_7 = x_8 + i x_9 =  f(\sigma) e^{i\omega\tau}.
$$
Here $\tau$ is the time and $0\le \sigma\le \pi$ is the world-sheet
parameter. $f(\sigma)$ is a profile that satisfies 
$f(0)=f(\pi) = 0$.

This configuration describes an open string that arches out
of the $0,1,2,3$ hyperplane of the D3-brane and into
the $6,7,8,9$ dimensions, with a profile $f(\sigma)$.
It is rotating in the $6,7,8,9$ dimensions with angular
frequency $\omega$.
It is stabilized by magnetic forces similarly to those studied
in \cite{Myers,MST} for a D-brane moving in an RR-field strength.
The magnetic force on a piece of string of length $\Delta l$
moving in an $H$-field with velocity $v$ is $H v \Delta l$ and
is perpendicular to the string and to $v$.

\vskip 0.5cm
\noindent\putfig{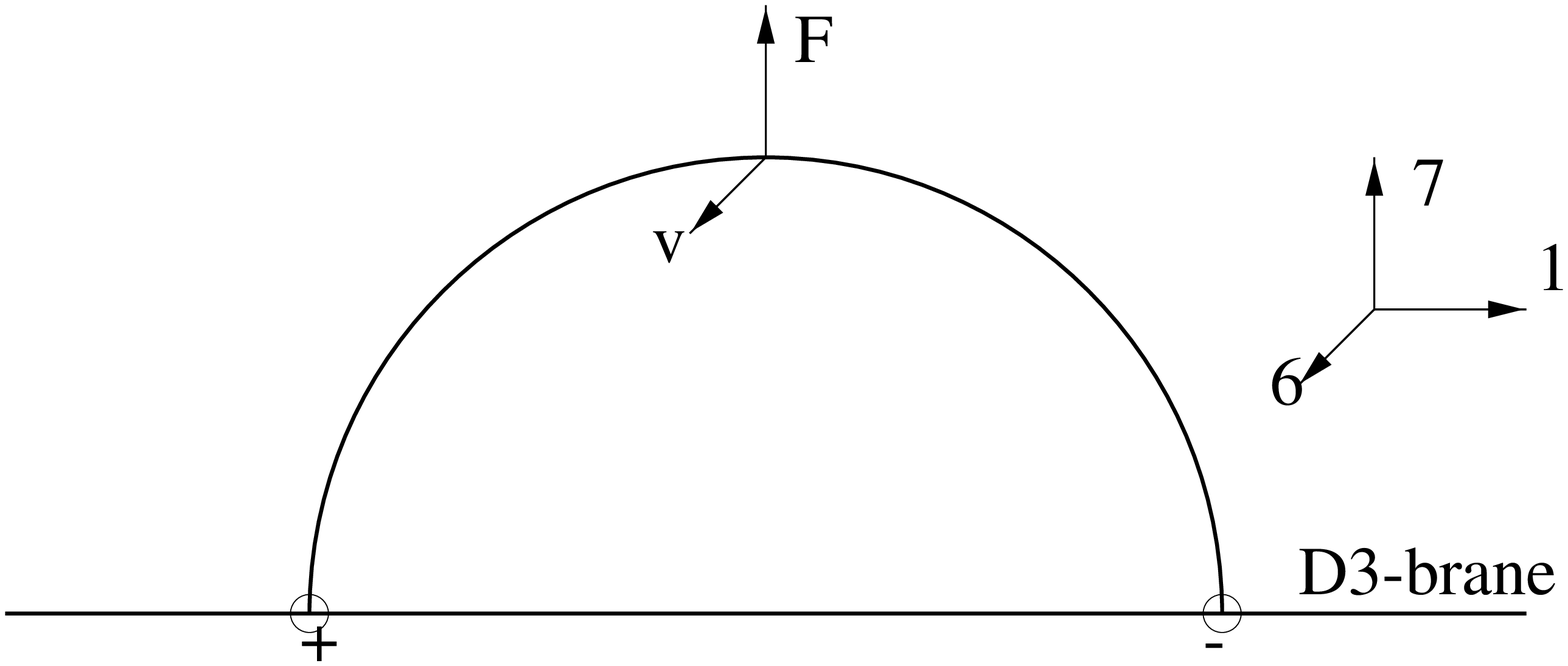}
{Fig.1: {The dipole on a D3-brane is a string that arches out
into the $6-7$ directions. The D3-brane is stretched along
the $1^{st}$ direction (and directions $2,3$ that are not shown)
and is at the origin of the $6-7$ plane. The generalized
magnetic force $F$ is perpendicular to the velocity and the string.}}{100mm}
\vskip 0.5cm

The string will be stable if the tension cancels the magnetic force.
Thus the profile is determined by the equation of balance of forces:
\be\label{eqprof}
\frac{\a' \frac{\pi^2}{L^2}f''}{1+\frac{\pi^2}{L^2}{f'}^2} 
  = H \omega f \sqrt{1+\frac{\pi^2}{L^2}{f'}^2}.
\ee
The angular momentum is then determined to be:
$$
J = \omega^2 \int_0^\pi f \sqrt{1+\frac{\pi^2}{L^2}{f'}^2} d\sigma.
$$
These equations are written in the non-relativistic approximation,
$\omega f \ll 1$,
in which case we can also neglect the centrifugal force.
The full relativistic equations can be derived from the world-sheet
action of the string in the background $H$-field, but 
we will not do that here.
We also assume that the string is not radiating gravitational
or $B$-field energy away. This can be justified by taking
the limit $b\rightarrow\infty$ and noting that in this case
there is a large rescaling  of the metric near the D-brane,
relative to the metric away from the brane, at infinity, by
a factor of $\sqrt{1+b^2}$ (see section (\ref{bps})).
Note that the non-relativistic approximation is also consistent with the
absence of radiation.

Equation (\ref{eqprof}) is a $2^{nd}$ order differential
equation for $f(\sigma)$ that should be solved subject to
the boundary conditions $f(0)=f(\pi) = 0$.
This boundary condition should, in principle determine 
$\omega$ as a function of $L$ and we could then calculate the
angular momentum, $J$ as a function of $L$.

Let us check how the profile $f(\sigma)$ behaves near
the ends. Looking for a solution of the form
 $f(\sigma)\sim \sigma^\delta$
we find $\delta =\frac{1}{3}$ so the string starts perpendicular
to the D3-brane, as is required for equilibrium.

Note that the above discussion assumes  that the angular momentum
is large (so that the classical equations of motion can be used).
We expect such objects to be heavy and of the order of magnitude
of the string scale. The dipoles of the dipole-theories are
light and can presumably be obtained by quantizing the open
strings in the background of the $H$ field.
%%% We hope to return to this subject in a later work.

\section{Behavior in a magnetic flux}\label{bps}
% =======================================================================
If we compactify a $U(1)$ 
dipole theory on $\MT{2}$ of area $A$ and put $n_m$
units of magnetic flux on the $\MT{2}$, the boundary
conditions for the dipole-fields acquire
extra phases. As a result, the lowest Kaluza-Klein
state of the dipoles has
a mass of $\frac{n_m |L|}{A}$, in the non-compact directions.
For an $SU(N)$ theory, we have to replace $n_m$ by $\frac{n_m}{N}$.
With no magnetic flux, the dipoles are massless.

We can test these statements by calculating the masses from BPS
arguments in the pinned-branes setting.
It is convenient to compactify on $\MT{6}$.
Let the radii of $\MT{6}$ be $R_1,\dots, R_6$.
The Taub-NUT (TN) space becomes a Kaluza-Klein soliton.
Let:
$$
M_{TN} \equiv \frac{1}{g_s^2} M_s^8 R_1\cdots R_5 R_6^2,\,\,
b \equiv M_s^{-2} B_{16},\,\,
M_{D3} \equiv \frac{N}{g_s} M_s^4 R_1 R_2 R_3,\,\,
M_{dp} \equiv k R_6^{-1}.
$$
We will eventually take the limit $R_1,\dots,R_5\rightarrow \infty$.
The BPS mass of a configuration of $N$ D3-branes with $k$ units
of momentum along $R_6^{-1}$ (that at the center of the Taub-NUT
space becomes $k$ units of R-symmetry charge) is:
\bear
m &=&
\sqrt{
(1+b^2) M_{TN}^2 + M_{D3}^2 + M_{dp}^2 
+2 \sqrt{ (1+b^2) M_{D3}^2 M_{TN}^2 + b^2 M_{TN}^2 M_{dp}^2}}
\nn\\ &\longrightarrow &
\sqrt{1+b^2} M_{TN} + M_{D3} 
+\frac{b^2 M_{dp}^2}{(1+b^2) M_{D3}} +\cdots
\nn
\eear
The arrow denotes  the limit $M_s R_1,\dots, M_s R_5\gg 1$.
Now, to turn on $n_m$ units of magnetic flux in direction $1,2$ we define:
$$
M_{D1} \equiv \frac{n_m}{g_s} M_s^2 R_3.
$$
The BPS mass formula, in the above limit, is (see \cite{CDGG}
for more details):
$$
m \rightarrow  \sqrt{1+b^2} M_{TN} + M_{D3}
 + \frac{4 b M_{D1} M_{dp}}{(1+b^2) M_{D3}} + \cdots
$$
where $(\cdots)$ are sub-leading corrections.
So, the mass of $k$ dipoles becomes:
$$
 \frac{4 b M_{D1} M_{dp}}{(1+b^2) M_{D3}} = 
  \frac{4 b k n_m}{N (1+b^2) M_s^2 R_1 R_2 R_6}.
$$
The dipole-length $L$ can be extracted from this formula by
comparing with the expected result of $\frac{n_m|L|}{A}$.
But before we do that we have to take into account the rescaling
of the metric due to the $B$-field. This rescaling is similar 
to the rescaling discussed in \cite{SWNCG} and can be 
found by calculating the BPS mass of Kaluza-Klein excitations
in the $1^{st}$ direction.
The result is (see \cite{CDGG} for details) $m = 1/\wR_1$ 
where $\wR_1 = \sqrt{1+b^2} R_1$. So the rescaling in the $1^{st}$
direction is by a factor of $\sqrt{1+b^2}$ and
we can identify the dipole length, $L$, as:
\be\label{bpsDL}
L = \frac{4 b}{\sqrt{1+b^2} M_s^2 R_6}.
\ee
Note that with the magnetic flux turned on, the
D3-brane becomes pinned to the center of
the Taub-NUT space, in general.
This follows from the fact that the transverse fluctuations,
described by the dipole fields, are massive.
There could, however, be special cases for which the D3-branes
are not pinned. This happens when $L$ is a rational fraction
of one of the radii of $\MT{2}$  and $n_m$ is a multiple of the
denominator of this fraction.

The behavior of the branes in a magnetic field that we discussed
above is a characteristic sign of the dipole-theories and we take it
as another evidence for the identification of the dipole-theories
with the low-energy description of the pinned branes.

From (\ref{bpsDL}) we see that even when $b\rightarrow\infty$
the dipole length $L$ is smaller than the string scale $M_s^{-1}$
unless we take $R_6\rightarrow 0$. This means that the T-dual picture
discussed at the end of section (\ref{dipole}) is better suited for
describing large dipole-lengths.
However, as we shall see in the next section, the generalization
to the $(2,0)$ is a different story!

\section{Generalization to the $(2,0)$ theory}\label{twoz}
% =======================================================================
There exists an interesting generalization of dipole-theories
to a deformation of the $(2,0)$ theory that depends on a tensor
$L^{\mu\nu}$. It can be similarly realized in the pinned-brane
setting by putting M5-branes (in directions
$1\dots 5$) inside a Taub-NUT space (that is homogeneous in 
directions $1\dots 6$ and the Taub-NUT circle is in the $7^{th}$ 
direction) and turning on a 3-form $C$-field that at infinity approaches
a constant $C_{127}$.
The analysis is similar to the one we have performed in this paper
and suggests that the theory on the pinned branes is parameterized
by the tensor $L^{12}$ that is proportional to $C_{127}$.

It is also amusing to conjecture that the theory has
disc-like or membrane-like excitations with the boundary
of the membrane in the $1-2$ plane and the area of the membrane being
proportional to $L^{12}$.
Those membranes are the generalization of the dipoles, but
unlike a dipole whose boundary is two disconnected points,
the boundary of an open membrane is a loop that can have a dynamics
of its own if the membrane is light.
We will call those objects ``{\bf discpoles}.''
The discpole-theory might be a simplified version of the noncommutative
$(2,0)$-theory \cite{ABS,NekSch,Berk} or OM-theory \cite{GMSS}.
(See also \cite{BerRoz} for related ideas.)

We can repeat the analysis of section (\ref{bps})
and compactify  this deformed $(2,0)$-theory on $\MT{3}$
with the analog of $n_m$ units of magnetic flux that is
a 3-form flux of the $(2,0)$-theory along $\MT{3}$.
Defining:
$$
M_{TN} \equiv M_p^9 R_1\cdots R_6 R_7^2,\,\,
M_{M5} \equiv N M_p^6 R_1\cdots R_6,\,\,
C\equiv M_p^3 C_{127},\,\,
M_{dp} \equiv k R_7^{-1},
$$
and define
$$
M_{M2} \equiv n_m M_p^3 R_4 R_5,
$$
we find that the mass of the discpole in the presence of flux is:
$$
\frac{4 k C n_m}{N (1+C^2) M_p^3 R_1 R_2 R_3 R_7}.
$$
The rescaling is now by a factor of $\sqrt{1+C^2}$ in both
the $1^{st}$ and $2^{nd}$ direction.
Therefore, the {\bf discpole-tensor} is given by:
$$
L^{12} = \frac{4 C}{M_p^3 R_7}.
$$
Now we see that in the limit $C\gg R_7\rightarrow\infty$
the scale of the discpole-theories can be kept below $M_p$.

Viewing these discpoles as a source term in eleven dimensional supergravity we
can, in principle, determine the detailed dynamics of the boundary.
%%%Results along that direction will be presented elsewhere.

\section*{Acknowledgments}
We are very grateful to J. Maldacena and N. Seiberg for helpful discussions
and comments. K.D. would like to thank the string theory group
at UPenn for stimulating discussions.
The research of K.D. is supported by Department of Energy grant No.
DE-FG02-90ER40542.
The research of O.J.G. was supported by National Science Foundation grant
No. PHY98-02484.
The research of G.R. is supported by NSF grant number NSF PHY-0070928 and by
a Helen and Martin Chooljian fellowship.

%%% ------------------------- CUT HERE ---------------------------------%
%%%%%%%%%%%%%%%%%%%%%%%%%%%%%%%%%%%%%%%%%%%%%%%%%%%%%%%%%%%%%%%%%%%%
%  B I B L I O G R A P H Y                                         %
%%%%%%%%%%%%%%%%%%%%%%%%%%%%%%%%%%%%%%%%%%%%%%%%%%%%%%%%%%%%%%%%%%%%
%%% \documentstyle[12pt]{article}
\def\np#1#2#3{{Nucl.\ Phys.} {\bf B#1} (#2) #3}
\def\pl#1#2#3{{Phys.\ Lett.} {\bf B#1} (#2) #3}
\def\physrev#1#2#3{{Phys.\ Rev.\ Lett.} {\bf #1} (#2) #3}
\def\prd#1#2#3{{Phys.\ Rev.} {\bf D#1} (#2) #3}
\def\ap#1#2#3{{\it Ann.\ Phys.} {\bf #1} (#2) #3}
\def\ppt#1#2#3{{Phys.\ Rep.} {\bf #1} (#2) #3}
\def\rmp#1#2#3{{\it Rev.\ Mod.\ Phys.} {\bf #1} (#2) #3}
\def\cmp#1#2#3{{\it Comm.\ Math.\ Phys.} {\bf #1} (#2) #3}
\def\mpla#1#2#3{{Mod.\ Phys.\ Lett.} {\bf #1} (#2) #3}
\def\jhep#1#2#3{{JHEP} {\bf #1}, #3 (#2)}
\def\atmp#1#2#3{{Adv.\ Theor.\ Math.\ Phys.} {\bf #1} (#2) #3}
\def\jgp#1#2#3{{\it J.\ Geom.\ Phys.} {\bf #1} (#2) #3}
\def\cqg#1#2#3{{\it Class.\ Quant.\ Grav.} {\bf #1} (#2) #3}
\def\hepth#1{{[hep-th/{#1}]}}

%%% \begin{document}

\end{document}